# Plume dynamics and nanoparticle formation in ultra-short laser ablation of metals

T. E. Itina[a]

*Hubert Curien Laboratory, UMR CNRS 5516/Lyon University, 18 rue du Prof. Benoit Lauras, Bat. F, 42000, Saint-*

*Etienne, France*

Nanoparticle formation by femtosecond laser ablation is investigated for different experimental conditions. Dynamics of the laser plume expansion is visualized based on numerical simulations and the possibility of primary nanoparticle formation is analyzed. Calculations are performed for metals under different background conditions. The effects of the background environment are considered, including initial expansion stage of the ejected atoms and clusters, nucleation, collisions, and longer scale nanoparticle evolution. Nanoparticle size distribution is shown to evolve from a decreasing function to the commonly observed bell-shaped distribution as early as at 50μs after the beginning of the laser pulse. However, at the end of the initial plume expansion stage, the supersaturation parameter can be sufficiently high for a much longer diffusion-driven nanoparticle growth to enter into play. In this case, the major nanoparticle formation process is based on the so-called catastrophic nucleation accompanied by collisional growth, which determines the final size distribution. The obtained calculation results explain numerous experimental findings and help to predict both nanoparticle plume evolution and changes in its size distribution.

[a] Electronic mail: tatiana.itina@univ-st-etienne.fr

Ultra-short laser ablation is a unique tool for material nanostructuring and for nanoparticle synthesis.[1,2,3,4,5] This method provides possibilities of a precise control over the laser processing. In particular, by using laser ablation, nanoparticles with rather small and well-controlled sizes were formed from different materials. That is why, during last decade, numerous experiments have been performed, aiming at the investigation of the mechanisms of both ultra-short laser ablation and of nanoparticle formation. Thus, in the case if femtosecond laser ablation, laser energy deposition was shown to induce an explosive ejection of a mixture of clusters, molecules and atoms,[6,7,8,9,10] rather than an equilibrium surface evaporation. Despite a large number of the experimental results, the theoretical understanding of the physical and chemical mechanisms leading to the formation of nanoparticles during femtosecond laser ablation is still lacking. In particular, long post-ejection stage was not enough investigated. Another open question concerns the changes in the nanoparticle size distribution.

To explain the experimentally observed processes, a number of analytical and numerical models have been proposed.[11,10,12,13,14] Two extreme cases of either very low or relatively high background pressure are considered in most theoretical studies of laser plume expansion. Laser plume expansion into vacuum was described as self-similar adiabatic one with condensation phenomenon.[15,16,17,18,19] In the case of a much higher background pressure, shock waves were shown to be produced during the plume expansion into background gases.[20] To describe plume expansion, a system of Navier-Stocks equations was solved providing a wealth of information about the first 1-2 microseconds of the plasma plume expansion. Gas-dynamical models are, however, invalid if the velocity distribution of the ablated particles deviates from the Maxwellian distribution. These models, furthermore, hardly describe diffusion of the plume species in the background environment, or inter-component mixing. As a result, they are inefficient at the delays longer than several microseconds after the beginning of the laser pulse. To solve this issue, it was proposed to switch the hydrodynamic calculations to the Direct Monte Carlo simulations[21] at 1-2 μs. To study femtosecond laser ablation, such approaches as molecular dynamics[10,22,23,24] (MD), hydrodynamics[25] (HD), and combinations with the direct simulation Monte Carlo method (DSMC) were also proposed.[21,26] These studies demonstrated that nanoparticles start to be formed at the very beginning of the plume formation. In particular, upon a typical femtosecond laser interaction, the target material is decomposed in a mixture of gas and particles. In the MD simulations, these effects appeared as a result of the calculated dynamics of the ensemble of the considered atoms based on the employed interaction potential.

Here, attention is focused at the formation of metallic nanoparticles that have found many promising applications due to their unique plasmonic and chemical properties. To increase nanoparticle yield and to better control over the particle formation, either an ambient gas or a liquid is used in most of the experimental set-ups. However, the roles of both plume dynamics and background environment in nanoparticle formation are still under discussion.

To examine ultra-short laser ablated plume dynamics and nanoparticle evolution under realistic experimental conditions and to account for the fact that the ablated plume contains several components, DSMC calculations of the plume dynamics are first performed in the presence of an inert background gas (Ar) with pressure $P$=300 Pa. The initial conditions are set based on the parameterization of the MD results obtained at a delay of 200 ps after the beginning of the laser pulse (100 fs, 800 nm). Figure 1 shows separately spatial density distributions of atoms and clusters for two different delays after the laser pulse. Here, larger clusters were initially at the back of the plume. The larger are the clusters the less pronounced is the stopping by the gas. Furthermore, diffusion-driven collisional nucleation starts at the plume front too at sufficiently long delays.

When ablation takes place in the presence of a background gas, or a liquid, two main stages can be considered (Figure 1). At the first stage, the ablated material expands until its front is compressed and part of it is scattered back.[12,21] After that, efficient mixing of both ablated and background species takes place. This second stage takes much longer time mostly defined by bi-component diffusion coefficients.

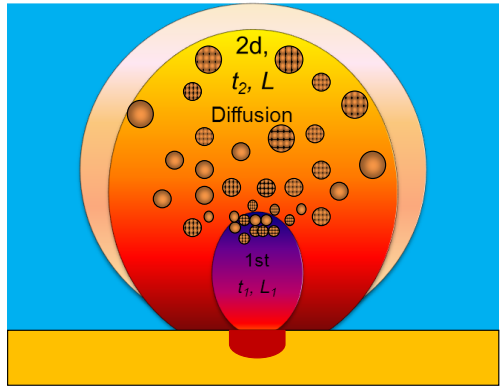

FIG. 1 (Color online) Schematics showing different stages of nanoparticle formation by laser ablation in the presence of a background environment. Here, 1- is plume expansion before its considerable deceleration by the background where rather small clusters are formed; 2 corresponds to the much longer diffusion of species in the background environment where collisional growth takes place.

Figure 2 demonstrates that, plume front starts experiencing a pronounced deceleration and practically stops at both plume- and gas-dependent delay (here, ~10μs). Theoretically, the initial expansion stage is described by a so-called snow-plow model, or by a blast-wave (or, shock-wave) model when shock waves are degenerated.

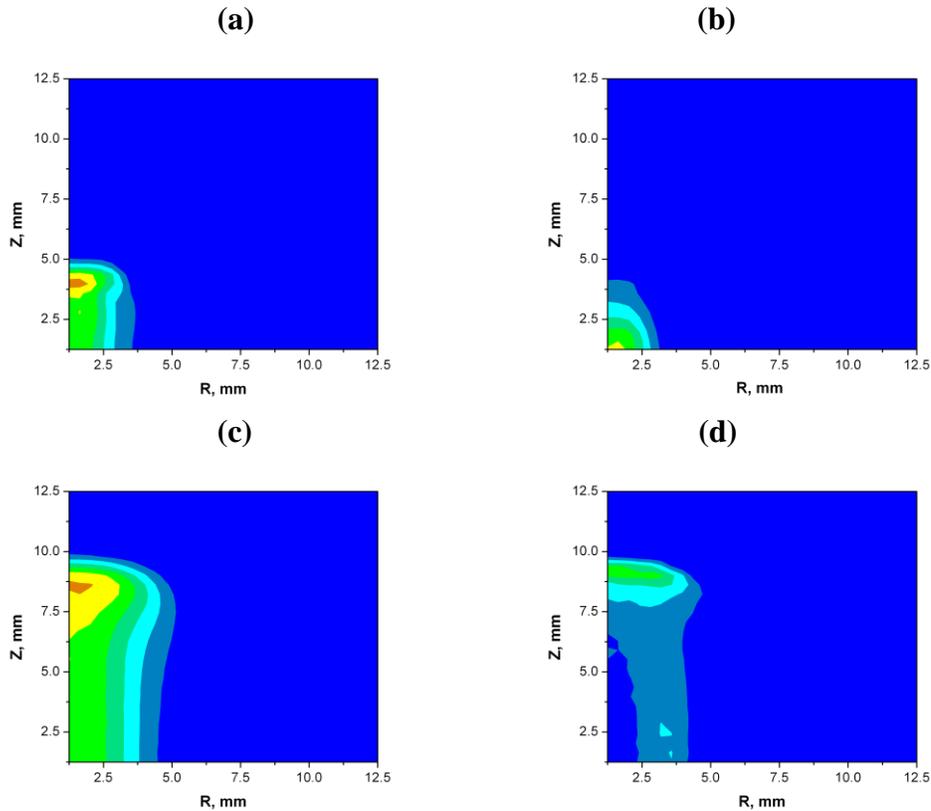

FIG. 2. (Color online) Calculated plume dynamics for Ni expansion in Ar gas at 300 Pa, (a) – density snapshot for atoms at $t$=0.55 μs, (b) –the same for clusters at $t$=0.55 μs; (c) –the same for atoms at t=10 μs, (d) –the same for clusters at $t$=10 μs

At the next stage, plume species get thermalized and a diffusion-driven regime enters into play.[21] In general, this effect starts when the mass of the adjacent background gas becomes comparable with the plume mass, or at a distance of[27]

$$L_1 = \sqrt[3]{\frac{3}{2\pi} \frac{MkT_b}{m_b}} P_b^{-1/3}, \qquad (1)$$

where $M$ is plume mass; $k$ is Boltzmann constant; $P_b$ is the background pressure; $T_b$ is temperature, and $m_b$ is the atomic weight of background gas species. In the considered case, plume contains atoms and clusters. Because clusters have larger mass, their stopping distance is longer, in agreement with the experiments of *Amoruso et al.*[1]

The corresponding nanoparticle size distributions are presented in Figure 3. One can see that after a sufficient delay, a peaked distribution appears instead of a decreasing function. This effect can be explained by collisional growth that is described by the general rate equation[28] having typically log-normal solutions[28,29]. It should be noted, however, that the amount of sufficiently large nanoparticles formed at such short delays is rather small and cannon explain the finally observed size distributions. The next much longer stage [Fig. 2(c,d)] includes plume mixing with the background followed by the rapid thermalization and a much more enhanced particle formation.

At distances shorter than $L_1$, metallic plume behaves almost as one-component gas so that its expansion can be described as an adiabatic process[30]. If nucleation occurs, a supersaturating condition $P > P_{eq}$ should be fulfilled, where the saturation pressure $P_{eq}$ is given by the Clausius–Clapeyron equation. As a result, the saturation ratio can be calculated as follows

$$\theta \approx 1 - \frac{T}{T_{eq}} = 1 - T\left\{\frac{1}{T_b} - \frac{k}{Q}\ln\left(\frac{P_{tot}}{P_b}\right)\right\}, \qquad (2)$$

where $T_{eq}$ is equilibrium temperature, $P_b$ is background pressure, and $P_{tot} = P$ is the metallic vapor pressure in the case of one-component adiabatic expansion; and $Q$ is the vaporization heat. These estimations are valid only until the delay determined by the beginning of the efficient mixing between the plume and ambient species ($t \sim 1\text{-}10$ μs).

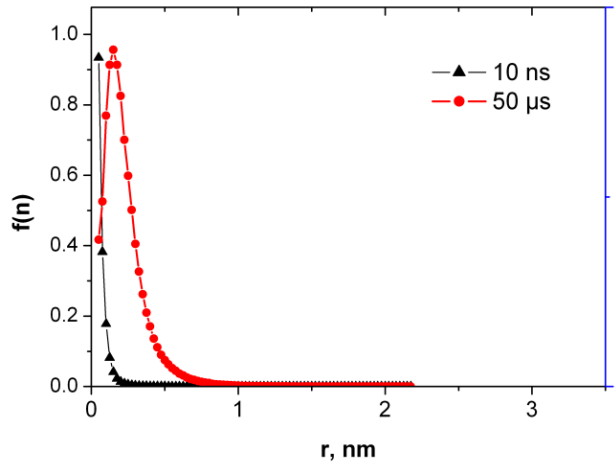

FIG. 3. (Color online) Size distributions calculated by using MD-DSMC model in the presence of 300 Pa of Ar at different time delays.

For the nucleation process to become considerable, saturation condition should be fulfilled. This means that pressure should exceed the saturated one, or plume temperature should drop down. Roughly, one can use the following estimation: $c_p T < Q$, where $C_p$ is the metal vapor heat capacity. This estimation gives, for instance, $T \sim 500$ K for Au and $\sim 670$ K for Ni.

During femtosecond ablation in the presence of a background gas, such plume temperatures are achieved at the plume thermalization stage. Upon both mixing and thermalization process, diffusion sets in, so that plume length can be approximately estimated as follows

$$L(t) = L_1 + \sqrt{D_i t}, \qquad (3)$$

where the $D_i$ is diffusion coefficient for different plume species $i$; and $t$ is time. At such delays, a considerable plume-background mixing occurs, so that nucleation mostly takes place in the two-component mixture, and $P_{tot}=P+P_b$. If the background pressure is high enough, this effect bursts nucleation, so that saturation is reached rather quickly. In addition, multiple laser pulses are typically applied in the ablation experiments. As a result, diffusion-driven nanoparticle growth can play an important role at rather long delays (stage 2 in Figure 1), that can be comparable to the inter-pulse time (typically, ~$10^{-3}$ s) at high background pressures.

For instance, in the presence of a sufficiently high-pressured background environment, such as atmospheric pressured gas or a liquid, diffusion-driven nucleation and aggregation processes start playing an important role at longer delays. The diffusion-driven nucleation leads to the formation of nucleus, whose size is controlled by the free energy as follows

$$\Delta G(n,c) = -nkT \ln(c/c_{eq}) + 4\pi a^2 n^{2/3} \sigma, \qquad (4)$$

where $k$ is the Boltzmann constant; $T$ is the temperature in Kelvins; $a$ is the effective radius; $c_{eq}$ is the equilibrium concentration of atoms/monomers; $n$ is the number of atoms/monomers in the nuclei; and $\sigma$ is the effective surface tension. The peak of the nucleation barrier corresponds to the critical cluster size

$$n_c = \left[\frac{8\pi a^2 \sigma}{3kT \ln(c/c_{eq})}\right]^3. \qquad (5)$$

The production rate of the supercritical nuclei is given by[31]

$$\nu(t) = K_c c^2 \exp\left[\frac{-\Delta G(n_c,c)}{kT}\right]. \qquad (6)$$

As a result, narrow size distributions are produced. The formed particles can collide and aggregate. These collisional processes are described by a simplified Smoluchowski equation[32]

$$\frac{dN}{dt} = K_{S-1} N_1 N_{S-1} - K_S N_1 N_S, \qquad (7)$$

where $s \geq 3$ is the number of monomers/primary particles in the particle, $N_s(t)$ is the time-dependent number density of the secondary particles containing s primary particles; $K_s = 4\pi(R_1 + R_s)(D_1 + D_S)$ is the attachment rate constant; $R_s = 1.2 r s^{1/3}$; $r$ is the average radius of the primary particle; $D_s = D_1 s^{-1/3}$ is the diffusion coefficient[31]. For simplicity, laser-induced fragmentation is not considered here. Several calculations are performed to study time-evolution of the size distribution by using equations (1-7) for multiple pulse cases. Because saturation is rather high, monomer radius here is as small as $a=1.59^{-10}$ and the radius of critical nuclei is recalculated at all the time-steps.

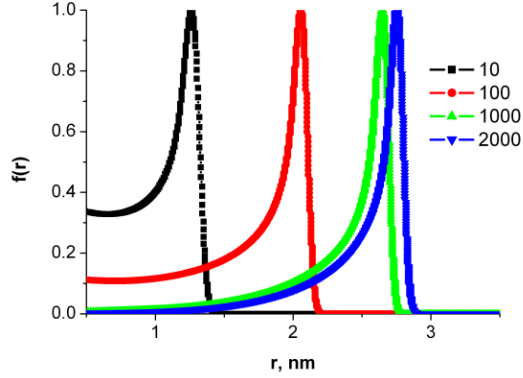

FIG. 4. Calculated size distribution obtained for 10, 100, 1000 and 2000 pulses. Here, laser frequency is 1 kHz, gold solution in water is considered with a=1.59$^{-10}$ m.

The obtained results (Figure 4) clearly show that when several pulses are applied these small nuclei grow for pulse number up to $10^3$ because tiny particles grow easier than larger ones. Here, collisions with atoms dominate in the growth process. The obtained results show that further increase in the number of pulses affects particle size distribution only slightly. This effect takes place if laser frequency is not too high, so that the ablated material has time to diffuse and concentration does not grow near the target. As a result, mean radius can remain rather small (in the nanometer range, smaller than ~3 nm here).

Finally, a criterion of the catastrophic nucleation[32] due to thermalization, diffusion and collisions of the ablated species in the background environment, is based on the inequality $n_c \leq 1$. This means that

$$n_c = \left[\frac{8\pi a^2 \sigma}{3kT \ln(c/c_{eq})}\right]^3 \sim 1. \qquad (8)$$

Therefore, if at the beginning of the second (Fig. 1), or diffusional expansion stage $\ln(S) = \ln(c/c_{eq}) = 8\pi a^2 \sigma / 3kT \geq 10$, this mechanism prevails in the nanoparticle formation. For instance, if $c \sim 10^{25}$ m$^{-3}$ and the equilibrium gold concentration in water[31] is $c_{eq} = 10^{15}$ m$^{-3}$, the condition (8) is satisfied.

In conclusion, both laser plume dynamics and nanoparticle formation have been investigated for femtosecond laser interactions in the presence of a background environment. We have considered several mechanisms of nanoparticle formation. The obtained calculation results have demonstrated the long time-evolution of plume species involves nucleation and growth determining the final size distribution that tends to a limiting one the increase in laser. Calculation results clearly explain several experimental observations including both longer time dynamics of nanoparticles and size distributions.[1,2,27] Furthermore, conditions are formulated for catastrophic nucleation to become the main mechanism of nanoparticle formation as a result of thermalization and collisions between among the species in the presence of a background environment.

The produced nanoparticles can be collected, form a colloid, or can be deposited at a substrate forming nanostructures. Therefore, the presented study is of interest for many applications where both metallic nanoparticles and nanostructures are used in nano-photonics, plasmonics, medicine, textile industry, and other promising fields.

**ACKNOWLEDGMENTS**

The research leading to these results received funding from the European Union Seventh Framework Programme (FP7/2007-2013) under Grant Agreement n° 280765 (BUONAPART-E). Partial support from the CNRS of France (PICS 6106 project) is gratefully acknowledged. Computer time was provided by CINES of France under the project x2015085015.